\documentclass[useAMS,usenatbib]{mn2e}
\usepackage{epsfig}
\usepackage{color} 
\usepackage{natbib}
\usepackage{deluxetable}
\usepackage{amssymb}
\usepackage{rotating}
\usepackage{graphicx}
\usepackage[colorlinks=true,citecolor=blue]{hyperref}

\title[Intranight optical variability of radio-quiet WLQs]
{Intranight Optical Variability of Radio-Quiet Weak Emission Line Quasars-III}
\author[Parveen Kumar,Gopal-Krishna and Hum Chand] {{Parveen
    Kumar$^{1}$\thanks{E-mail: parveen@aries.res.in (PK);
      gopaltani@gmail.com (GK); hum@aries.res.in (HC)},
    Gopal-Krishna$^{2}$, Hum Chand$^{1}$} \\ $^{1}$Aryabhatta Research
  Institute of Observational Sciences (ARIES), Manora Peak, Nainital,
  263002 India,\\ $^{2}$Centre for Excellence in Basic Sciences (CBS), University of Mumbai campus (Kalina), Mumbai 400098, India
}

\begin{document}
\date{Accepted ---. Received ---; in original form ---}

\pagerange{\pageref{firstpage}--\pageref{lastpage}} \pubyear{2014}

\maketitle

\label{firstpage}
\begin{abstract}
This is continuation of our programme to search for the elusive
radio-quiet BL Lacs, by carrying out a systematic search for
intranight optical variability (INOV) in a subset of `weak-line
quasars' which are already designated as `high-confidence BL Lac
candidate' and are also known to be radio-quiet. For $6$ such
radio-quiet weak-line quasars (RQWLQs), we present here new INOV
observations taken in 11 sessions of duration $>$3 hours each.
Combining these data with our previously published INOV monitoring of
RQWLQs in 19 sessions yields INOV observations for a set of 
  15 RQWLQs monitored in 30 sessions, each lasting more than 3
hours. The  30 differential light curves, thus obtained for the
15 RQWLQs, were subjected to a statistical analysis using the
F$-$test, and the deduced INOV characteristics of the RQWLQs then
compared with those published recently for several prominent AGN
classes, also applying the F$-$test. From our existing INOV
observations, there is a hint that RQWLQs in our sample show a
significantly higher INOV duty cycle than radio-quiet quasars and
radio lobe-dominated quasars. Two sessions when we have detected
strong (blazar-like) INOV for RQWLQs are pointed out, and these two
RQWLQs are therefore the best known candidates for radio-quiet BL
Lacs, deserving to be pursued. For a proper comparison with the INOV
properties already established for (brighter) members of several
prominent classes of AGN, a factor of $2-3$ improvement in the INOV
detection threshold for the RQWLQs is needed and it would be very
interesting to check if that would yield a significantly higher
estimate for INOV duty cycle than is found here.

\end{abstract}

\begin{keywords}
galaxies: active -- galaxies: photometry -- galaxies: jet -- quasars: general -- 
(galaxies:) BL Lacertae objects: general -- (galaxies:) quasars: emission lines
\end{keywords}


\section{Introduction} 

Flux variations at all energy bands is a well known characteristic of
active galactic nuclei (AGN)~\citep[e.g,
  see][]{Urry1995PASP..107..803U}. Optical flux variation on hour-like
time scale, which is commonly known as `intra-night optical
variability' (INOV), has emerged as a useful probe of
AGN~\citep{Wagner-Witzel1995ARA&A..33..163W,Ulrich1997ARA&A..35..445U,Wiita2006ASPC..350..183W}.
The past two decades have witnessed a large number of INOV studies
covering different classes of AGN, in order to study the physical
processes underlying this phenomenon occurring in the different AGN
classes~\citep{1989Natur.337..627M,Heidt1996A&A...305...42H,Carini-miller1990AJ....100..347C,
  Carini1991AJ....101.1196C,Carini1992AJ....104...15C,
  2007AJ....133..303C,Carini-Miller1992ApJ...385..146Co,
  1995MNRAS.274..701G,
  GopalKrishna1993A&A...271...89G,Stalin2004JApA...25....1S,Gupta2005A&A...440..855G,Joshi2011MNRAS.412.2717J,Joshi2013MNRAS.429.1717J,
  Goyal2013MNRAS.435.1300G, Chand2014MNRAS.441..726C,
  Diego2014arXiv1409.0468D}. These studies have led to theoretical
models for INOV~\citep[e.g, see
][]{Ulrich1997ARA&A..35..445U,Czerny2008MNRAS.386.1557C,Wiita2006ASPC..350..183W}.
For instance, in blazars, where pronounced INOV is observed, the cause
could be turbulence or localised particle acceleration events within
the non-thermal plasma flowing in a relativistic
jet~\citep[e.g.,][]{Wagner-Witzel1995ARA&A..33..163W,
  GopalKrishna2003ApJ...586L..25G,Singal-Gopal1985MNRAS.215..383S}. On
the other hand, in the case of radio-quiet quasars (RQQSOs) flares
occurring in the accretion disc might also play a significant if not
dominant, role in causing the
INOV~\citep{Manglam-Witta1993ApJ...406..420M}. In gamma-ray-loud
narrow-line Sefert1 galaxies, the detection of INOV on hour-like or
shorter time scale points to the presence of non-thermal jets with
large Doppler factors ~\citep{Paliya2013MNRAS.428.2450P}. Hence, INOV
studies of different classes of AGN can play a useful role in
improving the understanding of the AGN physics. \par

Weak emission line quasars (WLQs) is a relatively recently discovered
and rather enigmatic class of
AGN~\citep[e.g.,][]{Smith2007ApJ...663..118S,Plotkin2010AJ....139..390P,Heidt2011A&A...529A.162H}.
They exhibit abnormally weak broad emission-lines~\citep[i.e,
  rest-frame EW$ < 15.4$\AA~ for the Ly$+$NV emission-line
  complex,][]{DiamondStanic2009ApJ...699..782D}. The physical cause
for the abnormally weak line emission continues to be debated, as
summarized in the previous papers of this series ~\citep[][
  hereinafter Paper I \& Paper
  II]{Gopal2013MNRAS.430.1302G,Chand2014MNRAS.441..726C}. It may be
recalled that according to the currently prevailing view, the two
sub-classes of the most active AGN, called blazars, are BL Lac objects
(BLOs) and Highly-Polarized-Quasars (HPQs) which differ primarily in
the prominence of emission lines in the optical spectrum. But, whereas
HPQs have an abundant population of (usually weakly polarized)
radio-quiet counterparts (the RQQs), searches for radio-quiet analogs
of BLOs have so far remained unsuccessful, even probing the
radio-quiet subset of WLQs (RQWLQs) as possible candidates
~\citep[e.g.,][]{Jannuzi1993ApJ...404..100J,
  Londish2004MNRAS.352..903L}.

The explanations proposed for the WLQs basically fall in two
categories. One possible cause of the abnormality is the high mass of
the central BH ($M_{BH} > 3\times 10^9 M_{\sun}$) which can result in
an accretion disk too cold to emit strongly the ionizing UV photons,
even when its optical output is high
(~\citealt{Laor2011MNRAS.417..681L}; also,
~\citealt{Plotkin2010AJ....139..390P}). Alternatively, the covering
factor of the broad-line region (BLR) in WLQs could be at least an
order-of-magnitude smaller compared to the normal
QSOs~\citep[e.g.,][]{Nikolajuk2012MNRAS.420.2518N}. An extreme version
of this scenario is that in WLQs the accretion disk is relatively
recently established and hence a significant BLR is yet to
develop~\citep{Hryniewicz2010MNRAS.404.2028H,
  Liu2011ApJ...728L..44L}. Conceivably, a poor BLR could also result
from the weakness of the radiation pressure driven wind when the AGN
is operating at an exceptionally low accretion rate ($<~ 10^{-2}
\ \ to ~\ 10^{-3}\dot{M}_{Edd}$)
(~\citealt{Nicastro2003ApJ...589L..13N} ; also,
~\citealt{Elitzur2009ApJ...701L..91E}).

 While the above mechanisms may well operate commonly, a small
  fraction of RQWLQs may nonetheless turn out to be the radio-quiet
  counterparts of BL Lacs, such that the relativistic jet itself is
  radio-quiet. In order to pursue this interesting question, we
  started in 2012 an observational programme aimed at determining the
  INOV characteristics of RQWLQs (Papers I and II). In the present
  work (Paper III), we report the INOV results for $6$ of the RQWLQs
  which we monitored on $11$ nights. This paper is organized as
  follows. Section $2$ describes our RQWLQ sample. Observations and
data reduction procedures are described in Section $3$. Details of our
statistical analysis are presented in Section $4$, followed by a brief
discussion of the results in Section $5$. \par

\section{Observations and Data Reduction} 
\subsection{The sample of radio-quiet WLQs}

 Our sample for INOV monitoring (Table~\ref{tab:source_info}) was
  derived from the list of 86 radio-quiet WLQs published in Table 6
  of~\citet[]{Plotkin2010AJ....139..390P}, based on the SDSS Data
  Release 7~\citep[DR-7,][]{Abazajian2009ApJS..182..543A}. Out of that
  list, we included in our sample all $19$ objects which are brighter
  than R$\sim$18.5 and are classified therein as `high-confidence BL
  Lac candidate' (e.g., see Paper I and Paper II). Recently we have
  noticed that their classification as `high-confidence BL Lac
  candidate', is not fully secure, since it lacks a check for proper
  motion.

In fact, one of these $19$ WLQs, J090107.64+384658.8, has already been
argued to be galactic, based on its large proper motion of
$62.3\pm10.9$mas/yr~\citep{Wu2012ApJ...747...10W}. In view of this, we
have carried out a search for proper motion data for our set
of 19 WLQs, using the latest USNO
catalog~\citep{Monet2003AJ....125..984M} and the values are reproduced
in the last column of Table~\ref{tab:source_info}. From this table,
seven members of our set of 19 RQWLQs are seen to have a non-zero proper
motion. However, for three of them (viz. J110938$+$373611,
J140710$+$241853 and J161245$+$511817), the quoted proper motion is
not significant ($< 2.5\sigma$), making their galactic classification
uncertain (the choice of rms threshold is consistent with Londish et al. 2004). This is further corroborated by the fact that, based on a
multi-wavelength SED analysis, ~\cite{Wu2012ApJ...747...10W} have
confirmed extragalactic nature for the WLQ J110938$+$373611 for which
USNO proper motion is 10.8$\pm$4.5 mas/yr.
Likewise, for J140710$+$241853 and J161245$+$511817, non-zero
redshifts have been confirmed by
~\citet{Hewett2010MNRAS.405.2302H}. Therefore, we have retained these
three sources in our RQWLQ sample, and removed the remaining four
sources for which proper motion is detected above $2.5\sigma$ (these sources are
marked with ($\dagger$) in Table~\ref{tab:source_info}). Thus, the
proper motion check reduces our sample from 19 to 15
RQWLQs and these can be regarded as bona-fide `high-confidence BL Lac
candidates'. New observations of $6$ out of these 15 RQWLQs (marked by
asterisk ($*$) in Table~\ref{tab:source_info}) are reported in the
present work, based on 11 monitoring sessions. Note that one of the 4
excluded sources is J121929$+$471522 for which INOV detection with an
amplitude of of $\sim 7\%$ over a few hours was reported in Paper I.


\section{Observations and Data Reduction} 

\begin{table*}
\begin{minipage}{500mm}
{
\caption{The set of 19 `RQWLQs' initially selected for  our INOV programme$^a$. 
\label{tab:source_info}}
\begin{tabular}{lccc cc}
\hline
\multicolumn{1}{l}{IAU Name{\footnote {Result for the sources marked by $^{*}$ are reported in this paper. Although all these sources\\ are classified as `high-confidence BL Lac
candidate' in~\citet{Plotkin2010AJ....139..390P}, the $4$ sources \\ marked by $^{\dagger}$ are probably galactic, due to their significant proper motion.}}} &  R.A.(J2000) & Dec(J2000)                       &{\it R} &   $z$ & PM \\
         & (h m s)      &($ ^{\circ}$ $ ^{\prime}$ $ ^{\prime\prime }$) & (mag) & & mas/yr    \\
 (1)     &(2)             &(3)                             &(4)     &(5)  &(6) \\
\hline
\multicolumn{5}{l}{}\\

J081250.79$+$522531.05$^{*}$ &  08 12 50.80& $+$52 25 31 & 18.30 &1.152   & 00  \\
J084424.20$+$124546.00 & 08 44 24.20& $+$12 45 46 & 18.28 &2.466         & 00   \\
J090107.60$+$384659.00$^\dagger$ &   09 01 07.60& $+$38 46 59 & 18.21 &1.329       &  62.3$\pm$10.8  \\
J090843.25$+$285229.80$^{*}$ &   09 08 43.25& $+$28 52 29 & 18.55 &0.930  & 00  \\
J101353.45$+$492757.99 &   10 13 53.45& $+$49 27 57 & 18.40 &1.635       &  00 \\ 
J110938.50$+$373611.60 &   11 09 38.50& $+$37 36 11 & 18.72 &0.397       & 10.8$\pm$4.5  \\ 
J111401.31$+$222211.50$^\dagger$ &   11 14 01.31& $+$22 22 11 & 18.77 &2.121       &  10.2$\pm$2.0 \\ 
J115637.02$+$184856.50 &   11 56 37.02& $+$18 48 56 & 18.42 &1.956       & 00   \\
J121929.50$+$471522.00$^\dagger$ & 12 19 29.50& $+$47 15 22 & 17.66 &1.336    & 112.1$\pm$3.6  \\ 
J125219.50$+$264053.00 &   12 52 19.50& $+$26 40 53 & 17.94 &1.292       &  00  \\ 
J134601.29$+$585820.10$^{*}$ &  13 46 01.29& $+$58 58 20 & 17.73 &1.22    & 00  \\ 
J140710.26$+$241853.60$^{*}$ &  14 07 10.26& $+$24 18 53 & 18.49 &1.662   &  12.0$\pm$5.1  \\
J141200.04$+$634414.90$^{*}$ &  14 12 00.04& $+$63 44 14 & 17.97 &0.068   & 00  \\ 
J142943.60$+$385932.00 &   14 29 43.60& $+$38 59 32 & 17.56 &0.925       &  00  \\ 
J153044.10$+$231014.00 &   15 30 44.10& $+$23 10 14 & 17.32 &1.040       &  00  \\ 
J160410.22$+$432614.70$^{*}$ & 16 04 10.22& $+$43 26 14 & 18.04 &1.568    & 00   \\
J161245.68$+$511817.31 &   16 12 45.68& $+$51 18 17 & 17.70 &1.595       & 2.0$\pm$2.0 \\ 
J212416.05$-$074129.90 &   21 24 16.05& $+$07 41 29 & 18.29 &1.402       & 00 \\ 
J224749.56$+$134250.00$^\dagger$ &   22 47 49.56& $+$13 42 50 & 18.53 &1.179       &  14.1$\pm$3.6\\ 
                                                                               
\hline
\end{tabular}

}
 \end{minipage}
\end{table*}

\begin{table*}
\centering
\caption{Basic parameters and observing log for the $6$ RQWLQs and their chosen comparison stars (S1,S2).
\label{tab_cdq_comp}}
\begin{tabular}{ccc ccc c}\\
\hline

{IAU Name} &   Date       &   {R.A.(J2000)} & {Dec.(J2000)}                      & {\it g} & {\it r} & {\it g-r} \\
           &  dd.mm.yy    &   (h m s)       &($^\circ$ $^\prime$ $^{\prime\prime}$)   & (mag)   & (mag)   & (mag)     \\
{(1)}      & {(2)}        & {(3)}           & {(4)}                              & {(5)}   & {(6)}   & {(7)}     \\
\hline
\multicolumn{7}{l}{}\\

J081250.79$+$522531.0 &  01.01.2014      &08 12 50.79 &$+$52 25 31.0  &   18.30 &       18.05 &        0.25\\   
S1                    &                  &08 13 58.01 &$+$52 25 21.9  &   18.41 &       17.93 &        0.48\\
S2                    &                  &08 13 20.70 &$+$52 23 27.8  &   18.36 &       17.80 &        0.56\\
J081250.79$+$522531.0 &  02.01.2014      &08 12 50.79 &$+$52 25 31.0  &   18.30 &       18.05 &        0.25\\   
S1                    &                  &08 13 20.70 &$+$52 23 27.8  &   18.36 &       17.80 &        0.56\\
S2                    &                  &08 13 52.52 &$+$52 27 01.0  &   18.99 &       17.79 &        1.20\\
J090843.25$+$285229.8 &  01.02.2014      &09 08 43.25 &$+$28 52 29.8  &   18.55 &       18.50 &        0.05\\
S1                    &                  &09 09 00.07 &$+$28 56 48.4  &   19.24 &       18.22 &        1.02\\
S2                    &                  &09 08 58.83 &$+$28 55 38.9  &   18.93 &       17.93 &        1.00\\
J090843.25$+$285229.8 &  02.02.2014      &09 08 43.25 &$+$28 52 29.8  &   18.55 &       18.50 &        0.05\\
S1                    &                  &09 09 05.04 &$+$28 57 03.9  &   18.89 &       17.86 &        1.03\\
S2                    &                  &09 08 27.93 &$+$28 44 41.9  &   18.10 &       17.73 &        0.37\\
J090843.25$+$285229.8 &  01.04.2014      &09 08 43.25 &$+$28 52 29.8  &   18.55 &       18.50 &        0.05\\
S1                    &                  &09 08 32.43 &$+$28 50 38.5  &   18.85 &       18.07 &        0.78\\
S2                    &                  &09 08 41.76 &$+$28 52 17.5  &   19.31 &       18.85 &        1.46\\
J090843.25$+$285229.8 &  03.04.2014      &09 08 43.25 &$+$28 52 29.8  &   18.55 &       18.50 &        0.05\\
S1                    &                  &09 08 15.05 &$+$28 48 39.6  &   18.27 &       17.83 &        0.44\\
S2                    &                  &09 08 49.14 &$+$28 45 42.2  &   18.26 &       17.74 &        0.52\\
J134601.29$+$585820.1 &  01.04.2014      &13 46 01.29 &$+$58 58 20.1  &   17.96 &       17.74 &        0.22\\
 S1                   &                  &13 46 32.38 &$+$58 50 39.1  &   18.54 &       17.18 &        1.36\\
 S2                   &                  &13 45 46.99 &$+$59 01 59.5  &   17.65 &       17.12 &        0.53\\
J140710.26$+$241853.6 &  03.05.2014      &14 07 10.26 &$+$24 18 53.6  &   18.70 &       18.47 &        0.23\\
S1                    &                  &14 07 30.89 &$+$24 14 17.7  &   18.63 &       17.18 &        1.45\\
S2                    &                  &14 06 50.43 &$+$24 10 48.3  &   18.54 &       17.19 &        1.35\\
J141200.04$+$634414.9 &  04.05.2014      &14 12 00.04 &$+$63 44 14.9  &   17.77 &       17.05 &        0.72\\
S1                    &                  &14 12 03.97 &$+$63 43 05.9  &   17.63 &       17.11 &        0.52\\
S2                    &                  &14 12 35.80 &$+$63 37 16.1  &   17.53 &       16.94 &        0.59\\
J160410.22$+$432614.7 &  05.05.2014      &16 04 10.22 &$+$43 26 14.7  &   18.22 &       18.04 &        0.18\\
S1                    &                  &16 04 10.58 &$+$43 27 15.1  &   18.29 &       17.46 &        0.83\\
S2                    &                  &16 04 48.24 &$+$43 23 31.4  &   18.42 &       17.36 &        1.06\\
J160410.22$+$432614.7 &  30.05.2014      &16 04 10.22 &$+$43 26 14.7  &   18.22 &       18.04 &        0.18\\
S1                    &                  &16 04 15.97 &$+$43 19 17.7  &   18.75 &       17.66 &        1.09\\
S2                    &                  &16 04 48.24 &$+$43 23 31.4  &   18.42 &       17.36 &        1.06\\

\hline
\end{tabular}
\end{table*}


\begin{table*}
 \centering
 \begin{minipage}{500mm}
 {\small
 \caption{Observational details and INOV results for the set of 6 RQWLQs monitored in 11 sessions (present work).}
 \label{wl:tab_res}
 \begin{tabular}{@{}ccc cc rrr rrr ccc@{}}
 \hline  \multicolumn{1}{c}{RQWLQ} &{Date} &{T} &{N} 
 &\multicolumn{1}{c}{F-test values} 
 &\multicolumn{1}{c}{INOV status{\footnote{V=variable, i.e., confidence level
       $\ge 0.99$; PV=probable variable, i.e., $0.95-0.99$ confidence level;
       NV=non-variable,\\ i.e., confidence level $< 0.95$.
 Variability status inferred, $F^{\eta}$ values and INOV peak-to-peak amplitudes($\psi$) using the quasar-star1\\ and quasar-star2
 DLCs are separated by a comma.}}}
 &{$\sqrt { \langle \sigma^2_{i,err} \rangle}$}&{INOV amplitude}\\
 & dd.mm.yyyy& 
hr & &{$F_1^{\eta}$},{$F_2^{\eta}$}
 &F$_{\eta}$-test &(q-s) &$\psi_1(\%),\psi_2$(\%)&$\frac{}{}$\\
 (1)&(2) &(3) &(4) &(5)&(6)
 &(7) &(8)\\ \hline
 J081250.79$+$522531.0  &01.01.2014 &  3.59&    31&    0.43,    0.73&    NV, NV&   0.02&   3.43,       5.74\\         
 J081250.79$+$522531.0  &02.01.2014 &  3.46&    30&    0.38,    0.43&    NV, NV&   0.02&   2.98,       4.40\\         
 J090843.25$+$285229.8  &01.02.2014 &  4.69&    39&    0.46,    0.56&    NV, NV&   0.03&   8.29,       7.88\\         
 J090843.25$+$285229.8  &02.02.2014 &  4.29&    36&    0.62,    0.58&    NV, NV&   0.07&   30.39,     23.85\\         
 J090843.25$+$285229.8  &01.04.2014 &  4.92&    42&    0.50,    0.71&    NV, NV&   0.03&   6.23,       8.95\\         
 J090843.25$+$285229.8  &03.04.2014 &  4.30&    37&    0.58,    0.73&    NV, NV&   0.03&   10.82,     11.46\\         
 J134601.29$+$585820.1  &01.04.2014 &  4.21&    36&    0.48,    0.51&    NV, NV&   0.02&    4.18,      4.52\\         
 J140710.26$+$241853.6  &03.05.2014 &  3.00&    23&    4.19,    4.37&    V,  V&    0.05&    37.24,  36.73\\      
 J141200.04$+$634414.9  &04.05.2014 &  4.56&    37&    0.40,    0.26&    NV, NV&   0.04&    10.28,    5.75 \\         
 J160410.22$+$432614.7  &05.05.2014 &  4.62&    38&    1.01,    1.01&    NV, NV&   0.04&    26.05,    24.60\\         
 J160410.22$+$432614.7  &30.05.2014 &  4.32&    37&    1.08,    1.11&    NV, NV&   0.04&    17.61,    15.52\\         
                                                                                             
 \hline
 \end{tabular}
 }
 \end{minipage}
 \end{table*} 


 \begin{table*}
 \centering
 \begin{minipage}{500mm}
 {\small
 \caption{Observational details and INOV results for our entire set of 15 RQWLQs, covered in 30 monitoring sessions.}
 \label{wl3:tab_allthree}
 \begin{tabular}{@{}ccc cc rrr rrr ccc@{}}
 \hline  \multicolumn{1}{c}{RQWLQ} &{Date} &{T} &{N} 
 &\multicolumn{1}{c}{F-test values} 
 &\multicolumn{1}{c}{INOV status{\footnote{V=variable, i.e., confidence level
       $\ge 0.99$; PV=probable variable, i.e., $0.95-0.99$ confidence level;
       NV=non-variable,\\ i.e., confidence level $< 0.95$.
 Variability status, $F^{\eta}$ values and the INOV amplitudes($\psi$) derived using quasar-star1 and quasar-star2\\
 DLCs are separated by a comma.}}}
 &{$\sqrt { \langle \sigma^2_{i,err} \rangle}$}&{INOV amplitude}&{References}\\
 & dd.mm.yyyy& 
hr & &{$F_1^{\eta}$},{$F_2^{\eta}$}
 &F$_{\eta}$-test &&$\psi_1(\%),\psi_2$(\%)&$\frac{}{}$\\
 (1)&(2) &(3) &(4) &(5)&(6)
 &(7) &(8)&(9)\\ \hline
 J081250.79$+$522530.9  &23.01.2012 &  5.70&    13&    0.77,    0.59&   NV, NV&     0.01&    3.03,      1.94&  Paper I \\      
 J084424.24$+$124546.5  &26.02.2012 &  4.28&    17&    0.65,    0.63&   NV, NV&     0.02&    4.49,      3.49&  Paper I \\      
 J125219.47$+$264053.9  &25.02.2012 &  2.23&    09&    0.24,    0.37&   NV, NV&     0.01&    0.36,      1.39&  Paper I \\      
 J125219.47$+$264053.9  &23.03.2012 &  3.45&    09&    0.98,    1.02&   NV, NV&     0.02&    3.93,      3.87&  Paper I \\      
 J125219.47$+$264053.9  &19.05.2012 &  3.81&    15&    0.52,    0.54&   NV, NV&     0.02&    3.43,      3.76&  Paper I \\      
 J142943.64$+$385932.2  &27.02.2012 &  3.76&    18&    0.46,    1.41&   NV, NV&     0.01&    3.49,      4.58&  Paper I \\      
 J153044.07$+$231013.5  &27.04.2012 &  4.07&    20&    2.13,    1.48&   NV, NV&     0.01&    5.46,      3.81&  Paper I \\      
 J153044.07$+$231013.5  &19.05.2012 &  3.21&    13&    0.67,    0.58&   NV, NV&     0.02&    4.19,      4.02&  Paper I \\      
 J161245.68$+$511817.3  &18.05.2012 &  4.03&    16&    0.44,    0.44&   NV, NV&     0.03&    4.02,      3.87&  Paper I \\      
 J081250.79$+$522531.0  &12.11.2012 &  4.49&    50&    0.44,    0.78&   NV, NV&     0.04&   10.70,     12.91&  Paper II\\      
 J084424.24$+$124546.5  &13.11.2012 &  3.93&    25&    0.23,    0.33&   NV, NV&     0.04&    3.91,      5.66&  Paper II\\      
 J084424.24$+$124546.5  &04.11.2013 &  3.23&    38&    0.45,    0.50&   NV, NV&     0.02&    6.15,      6.95&  Paper II\\      
 J090843.25$+$285229.8  &09.02.2013 &  3.90&    32&    0.33,    0.44&   NV, NV&     0.04&    5.06,      8.90&  Paper II\\      
 J090843.25$+$285229.8  &10.02.2013 &  4.02&    33&    3.01,    3.14&   V,  V&      0.04&   31.73,     30.20&  Paper II\\      
 J101353.45$+$492757.9  &01.01.2014 &  4.43&    37&    1.96,    1.58&   PV, NV&     0.02&   12.79,     11.07&  Paper II\\      
 J101353.45$+$492757.9  &02.01.2014 &  4.58&    32&    1.10,    0.78&   NV, NV&     0.02&   10.68,      7.31&  Paper II\\      
 J110938.50$+$373611.6  &10.02.2013 &  4.43&    36&    0.54,    0.52&   NV, NV&     0.03&    9.89,      9.14&  Paper II\\      
 J115637.02$+$184856.5  &15.01.2013 &  5.05&    41&    0.59,    0.74&   NV, NV&     0.03&    7.60,      7.97&  Paper II\\      
 J212416.05$-$074129.9  &12.11.2012 &  3.40&    37&    1.08,    1.07&   NV, NV&     0.07&   33.33,     35.20&  Paper II\\      
 J081250.79$+$522531.0  &01.01.2014 &  3.59&    31&    0.43,    0.73&    NV, NV&     0.02&   3.43,       5.74&  Present work\\         
 J081250.79$+$522531.0  &02.01.2014 &  3.46&    30&    0.38,    0.43&    NV, NV&     0.02&   2.98,       4.40&  Present work\\         
 J090843.25$+$285229.8  &01.02.2014 &  4.69&    39&    0.46,    0.56&    NV, NV&     0.03&   8.29,       7.88&  Present work\\         
 J090843.25$+$285229.8  &02.02.2014 &  4.29&    36&    0.62,    0.58&    NV, NV&     0.07&   30.39,     23.85&  Present work\\         
 J090843.25$+$285229.8  &01.04.2014 &  4.92&    42&    0.50,    0.71&    NV, NV&     0.03&   6.23,       8.95&  Present work\\         
 J090843.25$+$285229.8  &03.04.2014 &  4.30&    37&    0.58,    0.73&    NV, NV&     0.03&   10.82,     11.46&  Present work\\         
 J134601.29$+$585820.1  &01.04.2014 &  4.21&    36&    0.48,    0.51&    NV, NV&     0.02&    4.18,      4.52&  Present work\\         
 J140710.26$+$241853.6  &03.05.2014 &  3.00&    23&    4.19,    4.37&     V,  V&      0.05&    37.24,  36.73&  Present work\\       
 J141200.04$+$634414.9  &04.05.2014 &  4.56&    37&    0.40,    0.26&    NV, NV&     0.04&    10.28,    5.75&  Present work\\         
 J160410.22$+$432614.7  &05.05.2014 &  4.62&    38&    1.01,    1.01&    NV, NV&     0.04&    26.05,    24.60&  Present work\\         
 J160410.22$+$432614.7  &30.05.2014 &  4.32&    37&    1.08,    1.11&    NV, NV&     0.04&    17.61,    15.52&  Present work\\         
 \hline
 \end{tabular}
 }
 \end{minipage}
 \end{table*} 

\subsection{Photometric Monitoring Observations}
The programme to determine the INOV properties of RQWLQs, initially
reported in Paper I, has been primarily carried out using the 1.3-m
Devasthal Fast Optical Telescope (DFOT) of the Aryabhatta Research
Institute of Observational Sciences (ARIES) located at Devasthal,
India~\citep{Sagar2011Csi...101...8.25}. We have also used the 1.04-m
Sampurnanand and IUCAA Girawali Observatory (IGO) telescopes for
optical monitoring of a few of these sources (Paper I). The entire
monitoring was done in the r band and each time a given RQWLQ was
monitored continuously for not less than 3.5 hours, except in case of J140710.26+241853.6 when the duration was a bit shorter (3.0 hours, Table~\ref{wl3:tab_allthree}).  DFOT is a fast beam (f/4)
optical telescope with a pointing accuracy better than $10$ arcsec
RMS. It is equipped with a 2K $\times$ 2K Peltier-cooled Andor CCD
camera having a pixel size of 13.5 micron and a plate scale of 0.54
arcsec per pixel. The CCD covers a field of view of 18 arcmin on the
sky and is read out with $31$ kHz and $1000$ kHz speeds, with the
corresponding system RMS noise of $2.5$, $7$ e- and a gain of $0.7$,
$2$ e-/Analog to Digital Unit (ADU), respectively. The CCD used in our
observations was cooled thermo-electrically to -$85$ degC. The
duration of each science frame was about $5-7$ minutes, yielding a
typical SNR above $25-30$. The FWHM of the seeing disk during our
observing was generally $\sim$ 2.5 arcsec.

In our sample selection process, care was taken to identify at least
two, but usually more, comparison stars on the CCD frame that were
within about 1 mag of the target RQWLQ. This allowed us to pin down
and discount any comparison stars which showed variability
during our observations, thus permitting a reliable differential
photometry of the RQWLQ monitored.

\subsection{Data Reduction}
\label{wl:sec_data}
The pre-processing work on the raw images (bias subtraction,
flat-fielding, cosmic-ray removal and trimming) was carried out using
the standard tasks in the Image Reduction and Analysis Facility
{\textsc IRAF} \footnote{\textsc {Image Reduction and Analysis
    Facility (http://iraf.noao.edu/) }}. The instrumental magnitudes
of the RQWLQs and their comparison stars in the image frames were
determined by aperture photometry technique
~\citep{1992ASPC...25..297S, 1987PASP...99..191S}, using the Dominion
Astronomical Observatory Photometry \textrm{II} (DAOPHOT
II)\footnote{\textsc {Dominion Astrophysical Observatory Photometry}}.

The aperture photometry was carried out for four values of aperture
radii, $1 \times$FWHM, $2 \times$ FWHM, $3 \times$ FWHM and $4 \times$
FWHM. Seeing disk radius (=FWHM/2) for each CCD frame was determined
by averaging over $5$ adequately bright stars present within each CCD
frame. While the photometric data using the different aperture radii
were found to be in good agreement, the best S/N was almost always
found with aperture radius of $2 \times$ FWHM. Hence, we adopted it
for our final analysis.

To derive the Differential Light Curves (DLCs) of a given
target RQWLQ, we selected two steady comparison stars present within
the CCD frames, on the basis of their proximity to the target source,
both in location and apparent magnitude. Coordinates of the
comparison stars selected for each RQLWQ are given in Table
~\ref{tab_cdq_comp}. The $g-r$ color difference for our `quasar-star'
and `star-star' pairs is always $< 1.5$, with a median value of $0.56$
(column 7, Table ~\ref{tab_cdq_comp}). Detailed analyses by
~\citet{Carini1992AJ....104...15C} and ~\citet{2004MNRAS.350..175S}
have shown that color difference of this magnitude should
produce a negligible effect on the DLCs as the atmospheric attenuation
varies over a monitoring session. \par

Since the selected comparison stars are non-varying, as judged from
the steadiness of the star-star DLCs, any sharp fluctuation confined
to a single data point was taken to arise from improper removal of
cosmic rays, or some unknown instrumental effect, and such outlier
data points (deviating by more than 3$\sigma$ from the mean) were
removed from the affected DLCs, by applying a mean clip algorithm.  In
practice, such outliers were quite rare and never exceeded two data
points for any DLC, as displayed in Figure \ref{fig:lurve}.


\section{STATISTICAL ANALYSIS OF DLCs} 
For checking the presence of INOV in a DLC,
C-statistic~\citep{1997AJ....114..565J} has been the most commonly
used test. Although the `one-way analysis of variance' (ANOVA) is the
most powerful test for this purpose, it requires a longer data train
than is usually present in the available
DLCs~\citep{Diego2010AJ....139.1269D}. In our analysis we have not
used the C-test since, as pointed out
by~\citet{Diego2010AJ....139.1269D}, the C-statistics, which is based
on ratio of standard deviations is not a reliable test for INOV. This
is because : (i) C is not a linear operator (ii) the commonly adopted
critical value (C = 2.576) is too
conservative~\citep{Diego2010AJ....139.1269D}. At the same time, the
ANOVA test was not found feasible since most of our DLCs contains no
more than $40$ data points. Therefore, we have based our
statistical analysis on the F-test which employs the ratio of
variances as, F $=
variance(observed)/variance(expected)$~\citep{Diego2010AJ....139.1269D},
with its two versions : (i) the standard \emph{F$-$test} (hereafter
$F^{\eta}-$test,~\citet{Goyal2012A&A...544A..37G}) and (ii) scaled
F-test (hereafter
$F^{\kappa}-$test,~\citet{Joshi2011MNRAS.412.2717J}).
$F^{\kappa}-$test is mainly used in cases when a large magnitude
difference is present between the target object and the available
comparison stars ~\citep{Joshi2011MNRAS.412.2717J}. Except in Paper I,
we have adopted $F^{\eta}-$test since for all our RQWLQs, we have got
comparison stars fairly close in apparent magnitude to the target
object. An additional advantage of employing the $F^{\eta}-$test is
that our results for RQWLQs can be readily compared with those
available in the recent literature for other AGN classes
~\citep[][hereafter
  GGWSS13]{Goyal2013MNRAS.435.1300G}\footnote{Recently, de Diego
  (2014) has introduced an improved version of the F-test, called
  enhanced F-test, which includes data for several comparison stars, in
  order to enhance the power and reliability of the F-test. Here we
  have limited to the F$^{\eta}$-test, in order to facilitate
  comparison with other AGN classes, as mentioned above. The results
  for our entire INOV dataset for RQWLQs, based on the enhanced
  F-test, will be presented elsewhere.}. A point worth emphasizing here
is that while applying the $F^{\eta}-$test, it is specially important
to use the correct rms errors on the photometric data points. It has
been found that the magnitude errors returned by the routines in the
data reduction softwares of DAOPHOT and IRAF, are normally
underestimated by a factor $\eta$ ranging between $1.3$ and $1.75$, as
shown in various studies~\citep[e.g.,][] {1995MNRAS.274..701G,
  1999MNRAS.309..803G, Sagar2004MNRAS.348..176S,
  Stalin2004JApA...25....1S, Bachev2005MNRAS.358..774B}.
Recently~\citet{Goyal2012A&A...544A..37G} estimated the best-fit value
of $\eta$ to be $1.5$. Following them, $F^{\eta}-$test can be
expressed as :
\begin{equation} 
 \label{eq.fetest}
F_{1}^{\eta} = \frac{\sigma^{2}_{(q-s1)}}
{ \eta^2 \langle \sigma_{q-s1}^2 \rangle}, \nonumber  \\
\hspace{0.2cm} F_{2}^{\eta} = \frac{\sigma^{2}_{(q-s2)}}
{ \eta^2 \langle \sigma_{q-s2}^2 \rangle},\nonumber  \\
\hspace{0.2cm} F_{s1-s2}^{\eta} = \frac{\sigma^{2}_{(s1-s2)}}
{ \eta^2 \langle \sigma_{s1-s2}^2 \rangle}
\end{equation}
where $\sigma^{2}_{(q-s1)}$, $\sigma^{2}_{(q-s2)}$ and
$\sigma^{2}_{(s1-s2)}$ are the variances of the `quasar-star1',
`quasar-star2' and `star1-star2' DLCs and $\langle \sigma_{q-s1}^2
\rangle=\sum_\mathbf{i=0}^{N}\sigma^2_{i,err}(q-s1)/N$, $\langle
\sigma_{q-s2}^2 \rangle$ and $\langle \sigma_{s1-s2}^2 \rangle$ are
the mean square (formal) rms errors of the individual data points in
the `quasar-star1', `quasar-star2' and `star1-star2' DLCs,
respectively. $\eta$ is the scaling factor and is taken to be $1.5$
from~\citet{Goyal2012A&A...544A..37G}, as mentioned above.

The $F^{\eta}$-test is applied by calculating the $F$ values using
Eq.~\ref{eq.fetest}, and then comparing them with the
critical $F$ value, $F^{(\alpha)}_{\nu_{qs},\nu_{ss}}$, where $\alpha$
is the significance level set for the test, and $\nu_{qs}$ and
$\nu_{ss}$ are the degrees of freedom for the ` quasar-star' and
'star-star' DLCs. Here, we set two significance levels, $\alpha=$ 0.01
and 0.05, which correspond to confidence levels of greater than 99 and
95 per cent, respectively. If $F$ is found to exceed the critical
value adopted, the null hypothesis (i.e., no variability) is discarded
to the corresponding level of confidence. Thus, we mark a RQWLQ as
\emph{variable} (`V') if F-value is found to be $\ge F_{c}(0.99)$ for
both its DLCs, which corresponds to a confidence level $\ge 99$ per
cent, \emph{non-variable} (`NV') if any one out of two DLCs is found
to have F-value $\le F_{c}(0.95)$. The remaining cases are designated
as \emph{probably variable} (`PV').

The inferred INOV status of the DLCs of each RQWLQ, relative to two
selected comparison stars, is presented in Table~\ref{wl:tab_res}. In
the first 4 columns, we list the name of the RQWLQ, date of its
monitoring, duration of monitoring and the number of data points (N)
in the DLC. The next two columns give the computed F values and the
corresponding INOV status of the two DLCs of the RQWLQ, as inferred
from the application of the $F^{\eta}-$test (see above). Column $7$
gives the photometric error $\sigma_{i,err}(q-s)$ averaged over the
data points in the `quasar$-$star' DLCs (i.e., mean value for q-s1 and
q-s2 DLCs), which typically lies between 0.02 and 0.07 mag (without
the $\eta$ scaling mentioned above). The last column gives the
{\it peak-to-peak} amplitude $\psi$ of INOV, as defined
by~\citet{Romero1999A&AS..135..477R} .

\begin{equation} 
\psi= \sqrt{({D_{max}}-{D_{min}})^2-2\sigma^2} 
\end{equation} 

with  $D_{min,max}$ = minimum (maximum)  values in the RQWLQ DLC and $\sigma^2$=
$\eta^2$$\langle\sigma^2_{q-s}\rangle$, where,
$\eta$ =1.5~\citep{Goyal2012A&A...544A..37G}.

\par
We have computed the INOV duty cycle (DC) for our RQWLQ sample
using the definition of~\citet{Romero1999A&AS..135..477R},
\begin{equation} 
DC = 100\frac{\sum_{i=1}^n N_i(1/\Delta t_i)}{\sum_{i=1}^n (1/\Delta t_i)} 
{\rm per cent} 
\label{eqno1} 
\end{equation} 
where $\Delta t_i = \Delta t_{i,obs}(1+z)^{-1}$ is duration of the
monitoring session of a RQWLQ on the $i^{th}$ night, corrected for its
cosmological redshift, $z$. Since the duration of the observing
session for a given RQWLQ differs from night to night, the computation
of \textit{DC} has been weighted by the actual monitoring duration $\Delta t_i$
on the $i^{th}$ night. $N_i$ was set equal to 1, if INOV was detected
(i.e., `V' for both DLCs on the night), otherwise $N_i$ was taken as $0$.


\section{Results and discussion}
\label{wl:sec_dis}
The present work together with Paper I \& Paper II allows us to determine
the INOV characteristics of RQWLQs using the entire set of 15
  bona-fide RQWLQs covered in our programme launched about two years ago.
This is the first investigation of the INOV properties of radio-quiet
weak-line quasars and is targeted on their subset classified in the literature as good
candidates for radio quiet BL Lacs. For the entire set we have got
$30$ DLCs which are continuous and have a duration exceeding 3.5 hours in all except one case where the duration is 3.0 hours (average duration of the 30 DLCs being
$4.2$ hours, see Table~\ref{wl3:tab_allthree} and section $3.1$). Our INOV results are
based on the $F^{\eta}$-test, which is not only more reliable in
comparison to other feasible tests (section 4), but also offers an
additional advantage in that our INOV results for the RQWLQs can be
directly compared with those reported in recent literature for other
prominent AGN classes (see below).

The INOV results reported in Papers I \& II were based on a set
  of $10$ RQWLQs with $19$ DLCs, yielding INOV duty cycle of 4 per
  cent. In this study, we have been able to significantly enlarge the
  INOV database as we now have $30$ DLCs covering our entire set of
  $15$ RQWLQs. The INOV duty cycle for the entire set is found
  to be $\sim$ 5 percent (using $F^{\eta}$-test). In order to
ascertain the effect of likely uncertainty in the adopted value of
$\eta$, we have repeated the computation of INOV duty cycle for the
 $30$ DLCs of RQWLQs, setting two extreme values for $\eta$
($=1.3$ and $1.75$) reported in the literature \citep[][and references
  therein]{Goyal2012A&A...544A..37G}.The INOV DCs computed for
  these extreme values of $\eta$ are still 5 per cent. Thus, the
  $F^{\eta}$-test is found to give a consistent result over the
  maximum plausible range in $\eta$. \par

It is interesting to compare our DC estimates for RQWLQs with those
recently reported by GGWSS13 for several prominent AGN classes, again
using the $F^{\eta}$-test with $\eta$ set equal to 1.5. INOV duty
cycle estimated in their study is: $\sim$ 10\%(6\%) for radio-quiet
quasars (RQQs), $\sim$ 18\%(11\%) for radio-intermediate quasars
(RIQs), $\sim$ 5\%(3\%) for radio lobe-dominated quasars (LDQs),
$\sim$ 17\%(10\%) for radio core-dominated quasars with low optical
polarization (LPCDQs) , $\sim$ 43\%(38\%) for radio core-dominated
quasars with high optical polarization (HPCDQs) and $\sim$ 45\%(32\%)
for BL Lac objects (BLOs) (The values inside parentheses refer to the
DLCs showing INOV amplitude $\psi > 3$\%). Thus, the duty cycle
  of strong INOV ($\psi > 3$\%) found here for RQWLQs is similar to
  those reported (with $\psi > 3$\%) for RQQs, RIQs, LDQs and LPCDQs,
  while HPCDQs and BLOs have distinctly higher duty cycle. However,
this comparison is not strictly valid, given the fact that in the
observations of all these other AGN classes (GGWSS13), an INOV detection
threshold ($\psi_{lim}$ ) of 1$-$2 per cent had typically been
achieved. Being 1$-$2 mag fainter, the INOV detection threshold
reached for the present set of RQWLQs is less deep ($\psi_{lim}$
$\approx$ 5 per cent, Table~\ref{wl3:tab_allthree}). Thus, while
making comparison with the above mentioned other AGN classes, our
present estimate of INOV duty cycle for RQWLQs ($\sim$ $5$ per cent)
may be treated as a lower limit. This cautionary remark is
  underscored by the fact that both events of INOV detection reported
  here (Table 4) are marked by extremely large amplitudes ($\psi \sim 30\%$ peak-to-peak,
  occurring on hour-like time scale), rivaling
  blazars in their highly active
  phases~\citep[e.g,][]{Sagar2004MNRAS.348..176S,GopalKrishna2011MNRAS.416..101G,Goyal2012A&A...544A..37G}.
  Clearly, it would be very interesting to check if a factor of
  $2-3$ improvement in $\psi_{lim}$ would reveal many more events of INOV among RQWLQs, yielding a statistically robust estimate for the duty cycle of strong INOV ($\psi > 3$\%) for RQWLQs, which is distinctly higher than the present estimate of $\sim 5\%$, perhaps even
  approaching the high values established for blazars. \par

\begin{figure*}
\centering
\epsfig{figure=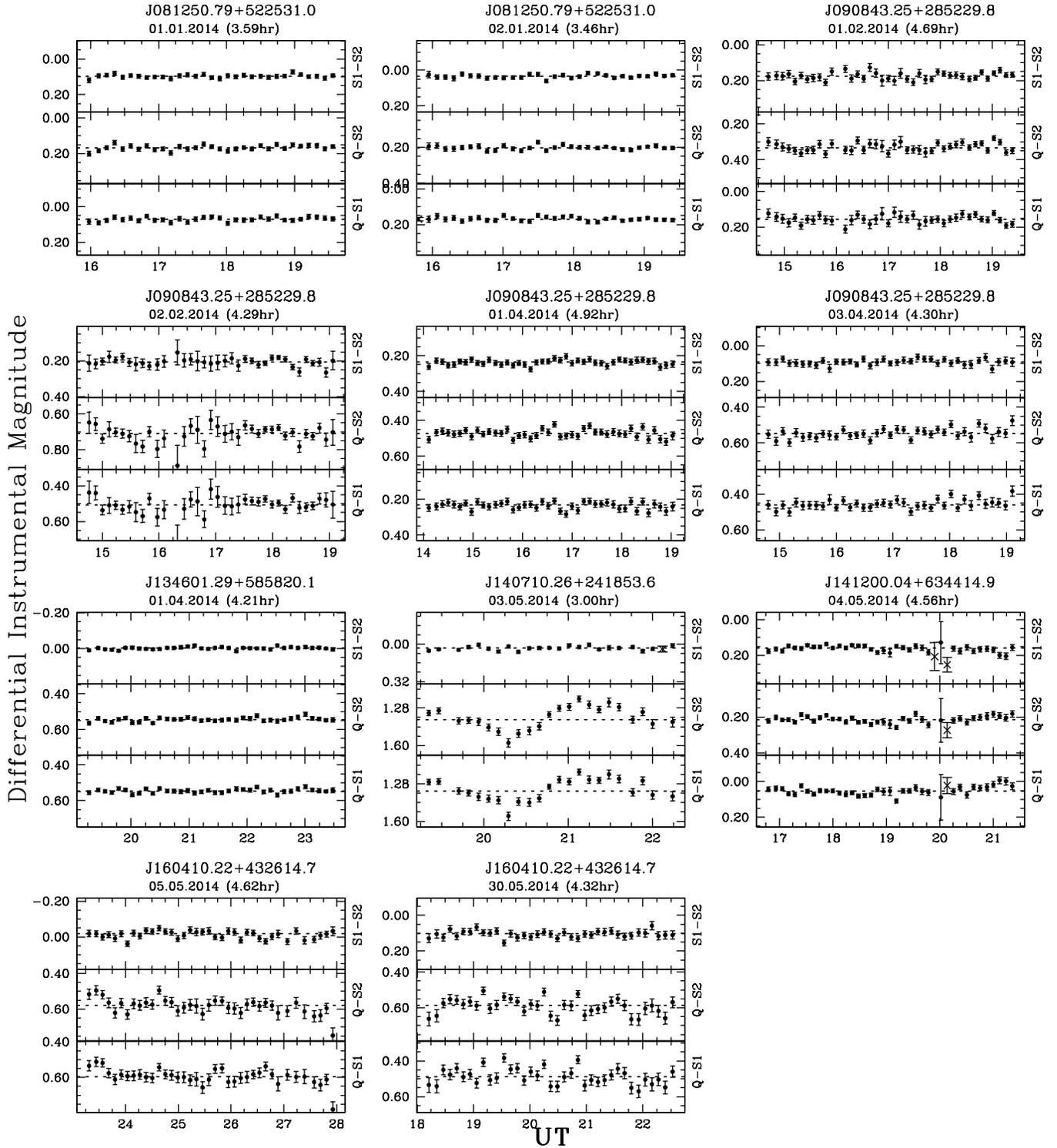}
\caption[]{Differential light curves (DLCs) for the $6$ RQWLQs from
  our sample. The name of the RQWLQ together with the date and
  duration of its monitoring are given at the top of each panel. In
  each panel the upper DLC is derived using the two non-varying
  comparison stars, while the lower two DLCs are the `quasar-star'
  DLCs, as defined in the labels on the right side. Apparently outlier
  point (at $> 3\sigma$) in the DLCs are marked with crosses and those
  points have been excluded from the statistical analysis.}
\label{fig:lurve}
 \end{figure*}

To summarize, the twin objectives pursued in our INOV study of RQWLQs
are (a) to find cases of very strong INOV ($\psi$ well above $3$ per cent), any such
RQWLQs would be outstanding candidates for the putative radio-quiet BL
Lacs, and (b) to quantify the INOV duty cycle for RQWLQs, in both
strong and weaker INOV regimes. With a significantly enlarged sample
of $30$ DLCs of RQWLQs in the present study, we now find that
their INOV duty cycle is about $5$ per cent, at a typical INOV
detection threshold of around $5$ per cent and a monitoring duration
of about 3$-$5 hours. In our programme, two of the RQWLQs were found
in two sessions to exhibit very strong INOV (amplitude $\psi 
>$ 10\%), a
level never  observed in our 2-decade long INOV
programme~\citep[summarized in
  GGWSS13,][]{Goyal2012A&A...544A..37G,Stalin2004JApA...25....1S},
except for BL Lacs and HPCDQs. The two RQWLQs, namely
  J090843.25$+$285229.8 ($\psi \sim 31$\% on 10.02.2013,
  Table~\ref{wl3:tab_allthree}) and J140710.26$+$241853.6 ( $\psi \sim
  36$\% on 03.05.2014, Figure \ref{fig:lurve},
  Table~\ref{wl3:tab_allthree}), are thus currently the best
available candidates for the elusive population of radio-quiet BL Lacs
and hence need to be followed up.
\section*{Acknowledgments}
 We thank an anonymous referee for the helpful suggestions. G-K thanks
 the National Academy of Sciences, India for the award of Platinum
 Jubilee Senior Scientist fellowship. \par

\bibliography{references}
\label{lastpage}
\end{document}